\documentclass[aps,prd,onecolumn,groupedaddress,showpacs,nofootinbib,amssymb]{revtex4}
\usepackage{graphicx,bm,color}
\usepackage[english]{babel}
\usepackage{amsmath}
\usepackage{amsfonts}

\begin{document}

\newcommand{\be}{\begin{equation}}
\newcommand{\ee}{\end{equation}}
\newcommand{\bea}{\begin{eqnarray}}
\newcommand{\eea}{\end{eqnarray}}
\newcommand{\nn}{\nonumber \\}
\newcommand{\e}{\mathrm{e}}

\title{Little Rip, $\Lambda$CDM and  singular dark energy cosmology from
Born-Infeld-$f(R)$ gravity}

\author{Andrey N. Makarenko$^{1,2}$,  Sergei D. Odintsov$^{3,4}$,  Gonzalo J. Olmo$^{5,6}$  \\ {\small
$^1$Tomsk State Pedagogical University, ul. Kievskaya, 60, 634061 Tomsk,
Russia} \\ {\small
$^2$National Research Tomsk State University, Lenin Avenue, 36, 634050 Tomsk,
Russia} \\
{\small $^3$Instituci\`{o} Catalana de Recerca i Estudis Avancats (ICREA),
Barcelona, Spain} \\
{\small
$^4$Institut de Ciencies de l Espai (CSIC-IEEC), Campus UAB,}\\ {\small Torre
C5-Par-2a- pl, E-08193 Bellaterra (Barcelona), Spain}\\
{\small $^5$Depto. de F\'{i}sica Te\'{o}rica \& IFIC , Universidad de Valencia - CSIC} \\ {\small  Burjassot 46100, Valencia, Spain }\\
{\small $^6$Depto. de F\'isica, Universidade Federal da
Para\'\i ba, 58051-900 Jo\~ao Pessoa, Para\'\i ba, Brazil}}

\begin{abstract}
 We study late-time cosmic accelerating dynamics from Born-Infeld-$f(R)$ gravity
in a simplified conformal approach.
We find that a variety of cosmic efects such as Little
Rip, $\Lambda$CDM universe and dark energy cosmology with finite-time future
singularities may occur. Unlike the convenient  Born-Infeld gravity
where in the absence of matter only de Sitter expansion may emerge,
apparentlly any FRW cosmology maybe reconstructed from this conformal version of the Born-Infeld-$f(R)$ theory. Despite the fact that the explicit form of $f(R)$ is fixed by the conformal ansatz, the relation between the two metrics in this approach
may be changed so as to bring out any desired FRW cosmology.
\end{abstract}

\hspace{13.1cm} IFIC/14-27

\maketitle

Modified gravity provides a very natural approach for the description of the evolution of the universe by generating  early-time as well as late-time cosmic acceleration via the modification of standard General Relativity (for recent reviews on modified gravity, see \cite{review}). It is remarkable that within this approach a unified description of the early-time inflation with late-time dark energy is possible, as first proved in ref.\cite{R}.

An interesting model of modified gravity extensively considered recently is the so-called Born-Infeld (BI) theory \cite{BI}, whose Palatini formulation avoids the appearance of ghosts. Some cosmological properties
of BI gravity have been discussed in a number of works \cite{cosmology}. 
The peculiarities of the Palatini formulation imply that the connection is compatible with an auxiliary metric algebraically related with the space-time metric via the matter sources. This type of algebraic relations make it  quite difficult to get consistent generalizations of the original BI model. However, we have recently demonstrated \cite{Eb6} that a non-perturbative and consistent generalization of BI gravity is possible when an $f(R)$ term is added to the original theory (hence dubbed BI-$f(R)$ theory). In the
present letter we consider the BI-$f(R)$ theory using a simplified conformal ansatz which, despite not being the most general approach (see ref.\cite{Eb6} for full details), it may be used to discuss a number of interesting situations. In fact, we explicitly construct Little Rip,  $\Lambda$CDM and  $\Lambda$CDM-like
cosmologies with finite-time future singularities. The possibility to reconstruct eventually arbitrary cosmologies like in $f(R)$ gravity\cite{LCDM} is briefly mentioned.

Let us briefly review the standard BI theory \cite{BI,Eb4}. The action for this theory is given by
\be
\label{e1}
S_{\text{EiBI}}=\frac{2}{\kappa}\int
d^4x\left[\sqrt{|\det{\left(g_{\mu\nu}+\kappa
R_{\mu\nu}(\Gamma)\right)}|}-\lambda\sqrt{|g|}\right]
+S_M[g,\Psi].
\ee
where $g_{\mu\nu}$ is the metric, $R_{\mu\nu}(\Gamma)={R^\alpha}_{\mu\alpha\nu}$ is the Ricci tensor, where 
\begin{equation}
{R^\alpha}_{\mu\beta\nu}=\partial_{\beta}
\Gamma^{\alpha}_{\nu\mu}-\partial_{\nu}
\Gamma^{\alpha}_{\mu\beta}+\Gamma^{\alpha}_{\beta\lambda}\Gamma^{\lambda}_{\nu\mu}-
\Gamma^{\alpha}_{\nu\lambda}\Gamma^{\lambda}_{\mu\beta}
\end{equation}
 is the Riemann tensor of the connection $ \Gamma_{\mu\nu}^{\lambda}$, which is {\it a priori} independent of the metric (Palatini formalism), 
and $\lambda$ is a dimensionless constant. The matter action depends on the matter fields, denoted generically by $\Psi$, and the metric $g_{\mu\nu}$ but not on the connection. The theory is considered under the Palatini formalism, i.e.,
$\Gamma^{\alpha}_{\mu\nu}$ is not assumed {\it a priori } to be the Levi-Civita connection of the metric
$g_{\mu\nu}$. Additionally, we assume that the Ricci tensor is symmetric and that there is no torsion.\\

Varying the action (\ref{e1}) with respect to $g_{\mu\nu}$ gives
\be
\label{e3}
\sqrt{q}\left(q^{-1}\right)^{\mu\nu}-\lambda \sqrt{g}g^{\mu\nu}=-\kappa
\sqrt{g} T^{\mu\nu}.
\ee
Here $T^{\mu\nu}$ is the standard energy-momentum tensor with indices raised
with the metric $g_{\mu\nu}$, $q=\det{q_{\mu\nu}}$,
\be
\label{eqq}
q_{\mu\nu}\equiv g_{\mu\nu}+\kappa R_{\mu\nu}(\Gamma).
\ee

Varying the action  (\ref{e1}) with respect to the connection we obtain 
\be
\label{e6}
\nabla_\alpha\left[\sqrt{q}\left(q^{-1}\right)^{\mu\nu}\right]=0 \ ,
\ee
where the covariant derivative is taken with respect to the independent connection. 
As is well-known, the covariant  derivative of a tensor density is given by
$$
\nabla_\mu \sqrt{q}=\partial_\mu \sqrt{q} - \Gamma^\alpha_{\mu\alpha}\sqrt{q}.
$$
For the vacuum case ($S_M=0$),
substituting  equation (\ref{e3}) into   (\ref{e6}) gives
\be
\nabla_\alpha\left[\sqrt{g}g^{\mu\nu}\right]=0 \ ,
\ee
which tells us that the connection is given by the Christofell symbols of the metric $g_{\mu\nu}$. That is
$q=\lambda g$ and we see that
$R_{\mu\nu}=\frac{\lambda -1}{\kappa}g_{\mu\nu}.$
If we consider the case $\lambda =1$ then
$$R_{\mu\nu}=0.$$
Thus, one sees the equivalence of BI gravity in vacuum with standard GR with cosmological
constant.

In the general case, equation (\ref{e6}) means that the tensor $q$ plays the role of an auxiliary
metric which is compatible with $\Gamma$
\be
\label{metric1}
\Gamma^\alpha_{\mu\nu}=\frac{1}{2} q^{\alpha\beta}\left(\partial_\mu
q_{\nu\beta}+\partial_\nu q_{\mu\beta}-\partial_\beta q_{\mu\nu}\right).
\ee

We now propose a modified action containing an arbitrary function $f(R)$, where
$R=g^{\mu\nu} R_{\mu\nu}(\Gamma)$. Such an action defines the BI-$F(R)$ family of gravity  theories and takes the form \cite{Eb6}:
\be
\label{act1}
S_{\text{EiBI}}=\frac{2}{\kappa}\int
d^4x\left[\sqrt{|\det{\left(g_{\mu\nu}+\kappa
R_{\mu\nu}(\Gamma)\right)}|}-\lambda\sqrt{|g|}\right]
+\int d^4x\sqrt{|g|}f(R)+S_M[g,\Gamma,\Psi].
\ee
  Varying the action  (\ref{act1}) with respect to the connection we obtain the
equation
\be
\label{equa1}
\nabla_\alpha\left[\sqrt{q}\left( q^{-1}\right)^{\mu\nu}+\sqrt{g} g^{\mu\nu}
f_R\right]=0 \ ,
\ee
where $f_R\equiv df/dR$. The corresponding equation obtained by variation over the metric has the form
\be
\label{e1_1}
\sqrt{q}\left(q^{-1}\right)^{\mu\nu}-\lambda
\sqrt{g}g^{\mu\nu}+\frac{\kappa}{2}\sqrt{g}g^{\mu\nu} f(R)-\kappa \sqrt{g}
f_R R^{\mu\nu}=-\kappa \sqrt{g}T^{\mu\nu}.
\ee
For simplicity, we now make the (simplifying) assumption that the tensor  $q_{\mu\nu}$ for the action
(\ref{act1}) is conformally proportional to the metric $g_{\mu\nu}$:
\be
\label{q1}
q_{\mu\nu}=k(t) g_{\mu\nu}.
\ee

In this case we have an auxiliary metric $u_{\mu\nu}$  which defines the
covariant derivative

\be
\label{metric2}
\Gamma^\alpha_{\mu\nu}=\frac{1}{2} u^{\alpha\beta}\left(\partial_\mu
u_{\nu\beta}+\partial_\nu u_{\mu\beta}-\partial_\beta u_{\mu\nu}\right).
\ee

Here
\be
\label{uq}
u_{\mu\nu}=(k(t)+f_R)g_{\mu\nu}.
\ee

For the condition (\ref{q1})  together with the definition $q_{\mu\nu}$ it is
clear that the Ricci tensor must also be proportional to the metric
$g_{\mu\nu}$. One can write the relationship between the Ricci tensor and the metric as
\be
\label{Ruq}
R_{\mu\nu}=\frac{1}{\kappa}(k(t)-1)g_{\mu\nu}.
\ee
Consider now the spatially-flat FRW universe with metric
\be
\label{FRW}
ds^{2}=-dt^{2}+a^{2}(t)(dx^{2}+dy^{2}+dz^{2})\  .
\ee
The auxiliary metric then takes the following form
\be \label{gg}
u_{\mu\nu}=u(t)\text{diag}(-1,a(t)^2,a(t)^2,a(t)^2) \ ,
\ee
where $u(t)=k(t)+f_R$. Suppose now that
$R_{\mu\nu}=r(t) g_{\mu\nu}$ where $r(t)$ is easy to find from the
Eq.(\ref{Ruq}). Now, the Christoffel symbols and Ricci
tensor of the metric (\ref{gg}) may be constructed leading to

\begin{eqnarray}
r(t)&=&\frac{3}{2} \left[2\frac{\ddot{a}}{a}+\frac{\dot{a}}{a}\frac{\dot{u}}{u}+\frac{\ddot{u}}{u}-\left(\frac{\dot{u}}{u}\right)^2\right] \\
r(t)&=&  \left[\frac{\ddot{a}}{a}+\frac{5}{2}\frac{\dot{a}}{a}\frac{\dot{u}}{u}+\frac{\ddot{u}}{2u}+2\left(\frac{\dot{a}}{a}\right)^2\right]  \ ,
\end{eqnarray}
where the upper dot denotes time derivative, i.e., $\dot{}\equiv \frac{d}{dt}$.
These two equations for $r(t)$ can be combined to get
\begin{eqnarray}\label{eq:Hubble-conf}
r(t)&=& 3\left(H+\frac{\dot{u}}{2u}\right)^2 \\ 
2\dot H&=& H \frac{\dot{u}}{u}+\frac{3}{2}\left(\frac{\dot{u}}{u}\right)^2-\frac{\ddot{u}}{u} \ ,
\end{eqnarray}
where we have defined $H$ as the Hubble rate ($H=\frac{\dot{a}}{a}$).
Using these two equations, one can verify that $\frac{\dot{u}}{u}=\frac{\dot{r}}{r}$, which leads to 
\be
\label{con1}
u(t)=c \ r(t)
\ee
where $c$ is a constant. The remaining equations lead us to
\be
\label{eeq1}
H=\pm\sqrt{\frac{u}{3c}}-\frac{\dot{u}}{2u}.
\ee
From this result, as shown in \cite{Eb6}, the form of the function $f(R)$ may be found explicitly
\be
\label{ff1}
f(R)=\frac{2}{\kappa}(\lambda-1)-R+\frac{c-\kappa}{8}R^2.
\ee
It should be noted that when we selected  the relation between the metric
$g_{\mu\nu}$ and the tensor $q_{\mu\nu}$ (\ref{q1}),
the form of the function $f(R)$ and the Eq.(\ref{eeq1}) is determined without
using  Eq.(\ref{e1_1}).
However,  Eq.(\ref{e1_1})  without  matter does not contradict the conditions
obtained previously and  allows us to fix some parameters.

The above consideration shows that this conformal approach allows to obtain the
solutions of equations of motion and the form of the function $f(R)$  only in
the absence of matter or matter with a constant energy-density $\rho=const$ and
$p=-\rho$. For more general forms of matter, the complete non-perturbative
formulation given in ref.\cite{Eb6} should be applied. Hence, we obtain the following form of $f(R)$
\be
\label{ff2}
f(R)=\frac{2}{\kappa}(\lambda-1)+2\rho-R+\frac{c-\kappa}{8}R^2.
\ee

Let us stress once more that the solutions in the presence of arbitrary matter may
be constructed, as shown in detail in ref.\cite{Eb6}.
All this leads to the solution of the system of Eqs.(\ref{equa1}) and
(\ref{e1_1}) with an arbitrary scale factor, which is determined from the
Eq.(\ref{eeq1}). From the other side, choosing the appropriate metric one finds
the relation between $g_{\mu\nu}$ and $q_{\mu\nu}$. Hence, the reconstruction
program appears to be completely different from the usual $f(R)$ gravity.
Let us consider several cosmologically-viable examples when the relation
between the metric  $g_{\mu\nu}$ and tensor $q_{\mu\nu}$ is explicitly chosen.

\subsection{Little Rip universe}
We can define the function $u$ as $u=u_0 e^{h\,t}$. Then from the
Eq.(\ref{eeq1}) it is easy to find the scale factor
\be \label{eqqq}
a= e^{\pm\frac{ 2 \sqrt{u_0}}{\sqrt{3c}  h}e^{\frac{h t}{2}}-\frac{h t}{2}},
\ee
and
\be \label{pl1}
H=-\frac{h}{2}\pm\frac{\sqrt{u_0}}{\sqrt{3} \sqrt{c}}e^{\frac{h\,t}{2}}.\ee

The effective equation of state (EoS) parameter becomes:
$$w_{eff}=-1-\frac{2\dot{H}}{3H^2}=
-1\mp\frac{4 h\sqrt{3} \sqrt{u_0 c}}{\left(3 \sqrt{c} h-2 \sqrt{3} e^{\frac{h
t}{2}} \sqrt{u_0}\right)^2} e^{\frac{h t}{2}}.$$
where the time-dependence  of the effective EoS is presented in Fig.1 for $h<0$.
The behavior of the effective  EoS for $h> 0$ is exactly the same as in Fig. 1, only the red line will meet the ''+'' sign in the equation  (\ref{pl1}) and blue -- ''-''.
\begin{figure}[tbp]
\begin{minipage}[h]{0.4\linewidth}
\includegraphics[width=7cm,height=5cm]{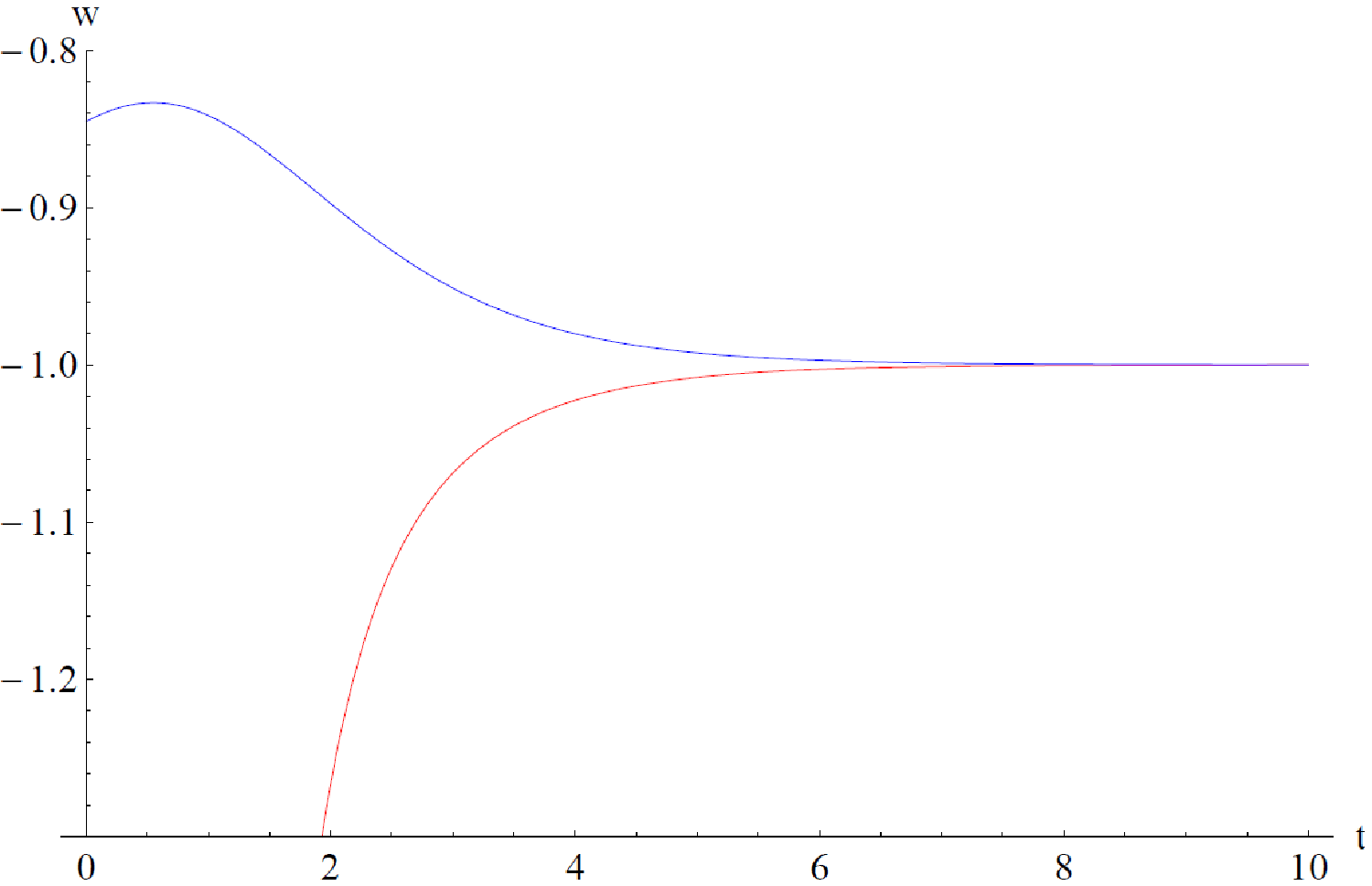}
\caption{$w_{eff}$ as a function of $t$ (for $u=e^{h\,t}$)
with the parameters
$c=0.1$, $h=-2$,
$u_0=3$ (blue line corresponds the  sign ''+'' in (\ref{eqqq})  and red -- "-" ).}
\end{minipage}
\hspace{10mm}
\begin{minipage}[h]{0.4\linewidth}
\includegraphics[width=7cm,height=5cm]{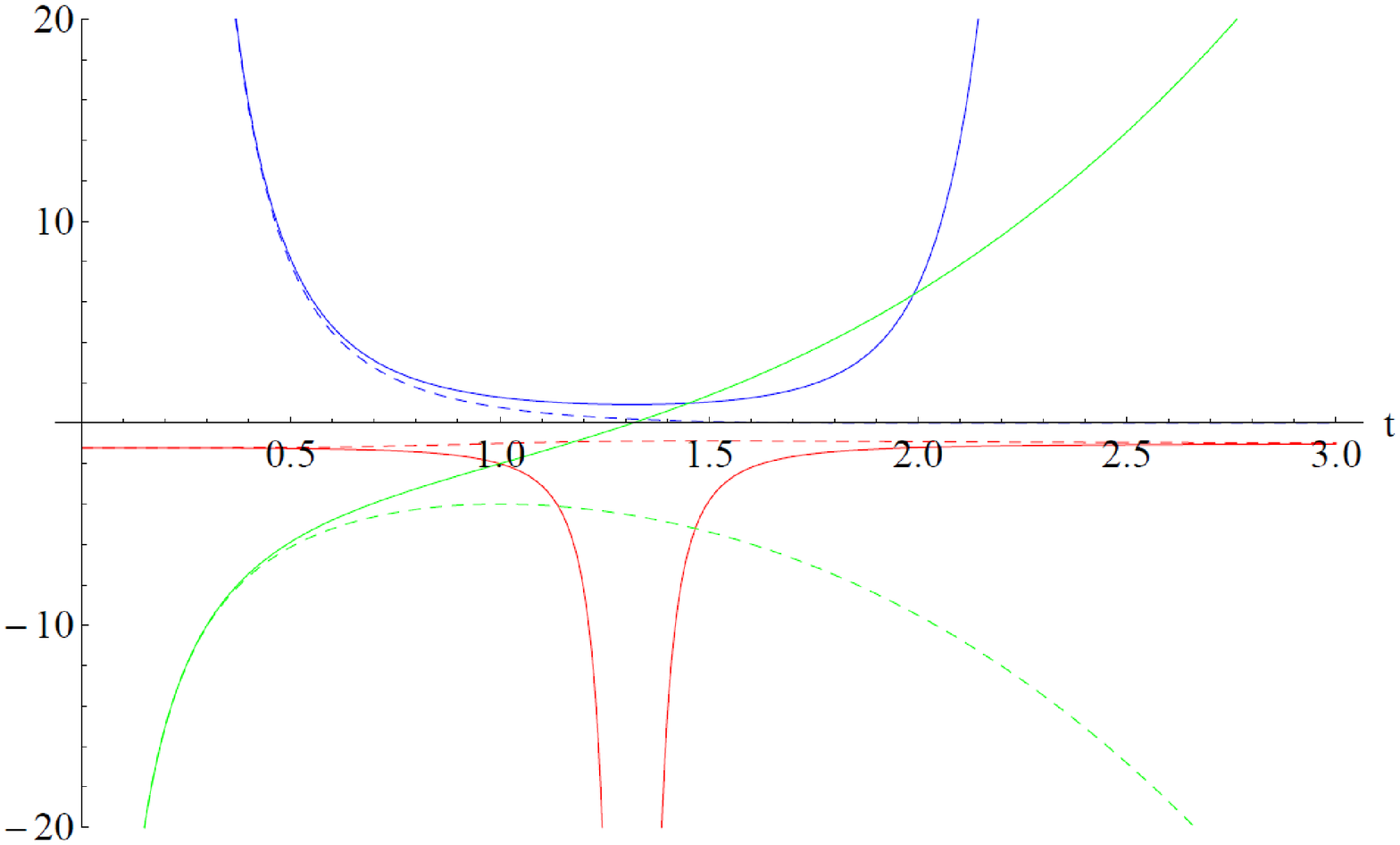}
\caption{$a$ (blue line), $H$ (green line) and $w_{eff}$ (red line) as a function of $t$  ($u=t^{h}$)
with the parameters
$c=1$, $h=6$,
$u_0=3$ (continuous line corresponds the sign ''+'' in (\ref{pl2})  and dashed -- "-" ).}
\end{minipage}
\label{graph_1}
\end{figure}

It is easy to see that for a  positive  $h$ (and sign ''+''  in the expression
(\ref{eqqq}))  we obtain the so-called Little Rip universe \cite{frampton11}
($a\to \infty$ and $H\to\infty$ at future infinity). For negative values of
$h$, the model  behaves as de Sitter universe at infinity. If in the expression
(\ref{eqqq}) we choose  the  sign ''+'' then  we  observe the exponentially
expanding universe.  If we select the sign ''-'' then the early universe   is
compressed and then it
expands, that is, we have the so-called bounce cosmology \cite{B-R-BCS}.
  The case with a positive value of the parameter $h$ and  sign ''-'' in the
Eq.(\ref{eqqq}) is not interesting, since the scale factor  tends to zero.
From Fig.1   it is clear that
  phantom-like ($w<-1$) as well as  quintessence-like ($-1/3<w<-1$) dark energy
cosmologies maybe obtained as solutions from our BI-$f(R)$ theory.

For the Little Rip scenario, the effective EoS is always less than $-1$, so
that the dark energy density increases with time,  but $w$ approaches $-1$
asymptotically and sufficiently rapidly that a singularity is avoided. But it
leads to a dissolution of bound structures at some point in the future (similar
to the effect of a Big Rip singularity).
As the universe expands, the relative acceleration between two points separated
by a distance $l$
is given by $l\ddot{a}/a$. If there is a particle with mass $m$ at each of
these points, an observer at one of the masses will measure an inertial force
on the other mass, as  \cite{frampton11}
\be
\label{FIN1}
F_{iner}=m\,l\,\ddot{a}/a=m\, l\left(\dot{H}+H^2\right).
\ee

Let us assume the two particles are bound by a constant force $F$ . If
$F_{iner}$ is greater than $F$, the two
particles become unbound. This is the Rip produced by the acelerating
expansion. We see that this situation will be realized if $H$ or/and $\dot{H}$
tends to infinity even in the BI-$f(R)$ gravity under consideration.
Indeed, we see that for $h$ positive
we receive the force $F_{iner}$ exponentially growing with time which tends to
infinity. For example, if $c=1$, $h=2$,
$u_0=1$ the disintegration of the Solar System  occurs at $F_{iner} \sim
10^{23}$, which corresponds to $t\approx 24.4$ Gyr.

Note that in the model under consideration the curvature is determined by the
metric $u_{\mu\nu}=u_0 e^{h\,t} g_{\mu\nu}$, and $R$ is equal to $\frac{4
e^{h\, t} u_0}{c}$ rather than $3 h^2-\frac{3 \sqrt{3} e^{\frac h\, t}{2} h
\sqrt{u_0}}{\sqrt{c}}+\frac{4 e^{h\, t}u_0}{c}$ as it should be for the metric
(\ref{eqqq}).

\subsection{ Power-law evolution}

For $u=u_0 t^h$ one finds
\be \label{pl2}
a=a_0 e^{\frac{\pm 2 t ^{1+\frac{h}{2}}\sqrt{u_0}}{\sqrt{3c} (2+h)}} t^{-h/2} ,
\ee
or
\be  \label{pl3}
a=a_0 e^{\frac{\pm 2 (t_0-t)^{1+\frac{h}{2}}\sqrt{u_0}}{\sqrt{3c} (h-2)}}
(t_0-t)^{-h/2} ,
\ee  for $u=u_0 (t_0-t)^{-h}$.
The Hubble parameter takes the form
$$
H=-\frac{h}{2 t}\pm\frac{t^{h/2} \sqrt{u_0}}{\sqrt{3} \sqrt{c}},$$
while for  $u=u_0 (t_0-t)^{h}$ one should  replace $t \to (t_0-t)$ and $h\to
-h$.
In the first case, the  effective EoS parameter  takes the form
$$w_{eff}=\frac{-3 c h (4+3 h)\pm 8 \sqrt{3 u_0 c}  h t^{1+\frac{h}{2}}-12
t^{2+h} u_0}{\left(3 \sqrt{c} h-2 \sqrt{3} t^{1+\frac{h}{2}}
\sqrt{u_0}\right)^2},$$
and we can build time-dependence of the EoS parameter
(Fig.2). We see that the effective EoS parameter approaches to minus one. The behavior
of the scale factor and the Hubble parameter can be illustrated by Fig. 2.

We see that the expanding universe corresponds the scale factor $a$ appropriate
to the plus sign in Eq.(\ref{pl2}).
This is again Little Rip  where the singularity is moved to infinity.
  In addition, we again observe a bouncing cosmology.
For $u=u_0 (t_0-t)^{-h}$  one gets

$$w_{eff}=\frac{-3 c h (-4+3 h) (t_0-t)^h\pm8 \sqrt{3} \sqrt{c} h
(t_0-t)^{1+\frac{h}{2}} \sqrt{u_0}-12 (t_0-t)^2 u_0}{\left(3 \sqrt{c}
h(t_0-t)^{h/2}-2 \sqrt{3} (t_0-t) \sqrt{u_0}\right)^2}.$$

This model  behaves almost like the model built with the metric (\ref{pl2}), but at the
moment $t=t_0$ we have a Big Rip singularity \cite{sing}. It should
be noted that in this case the effective EoS parameter  at this time is equal
to minus one.
The time-dependence of the EoS parameter is drawn in Fig.3.

\begin{figure}[tbp]
\begin{minipage}[h]{0.4\linewidth}
\includegraphics[width=7cm,height=5cm]{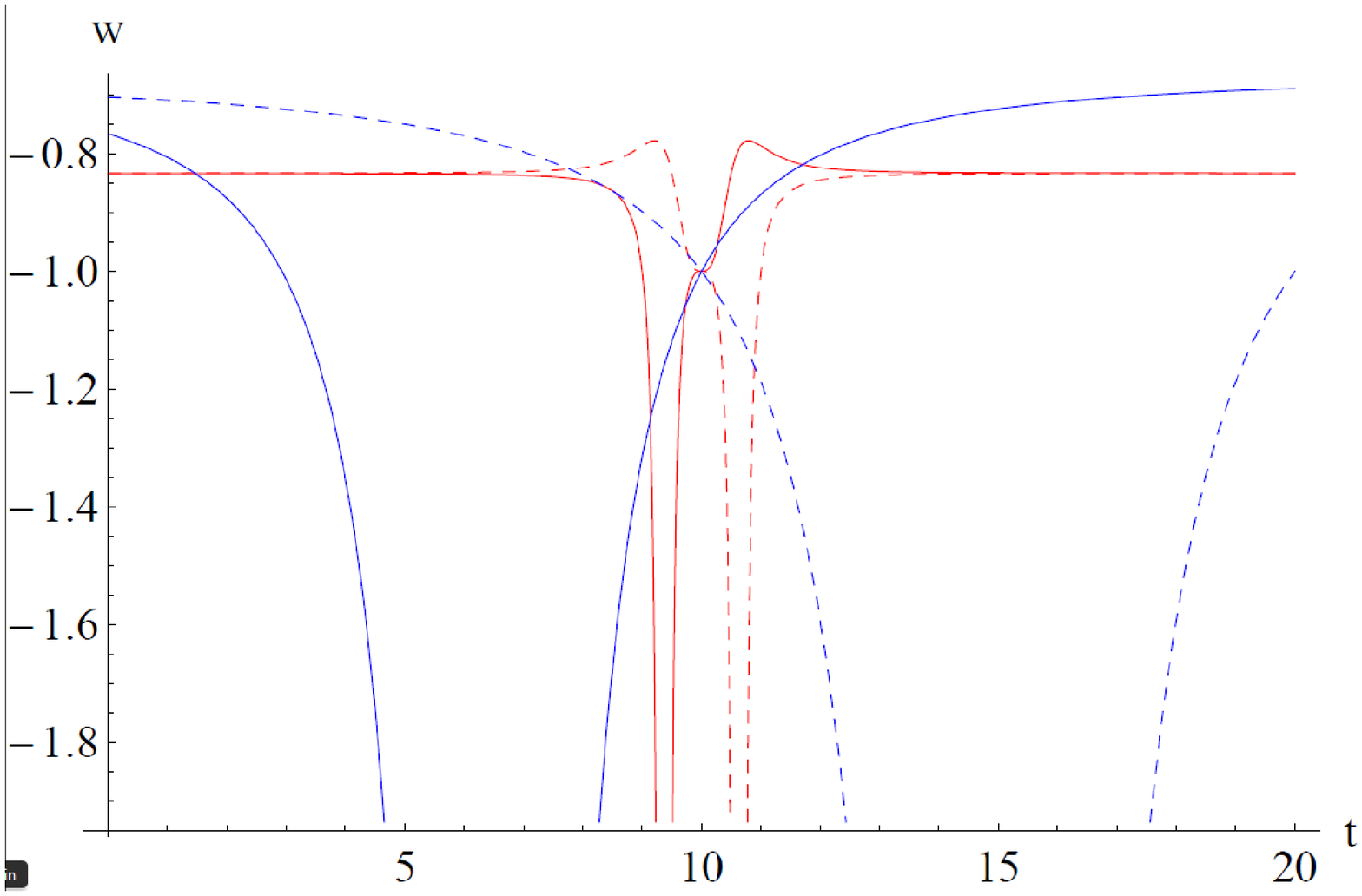}
\caption{$w_{eff}$ as a function of $t$ ($u=u_0 (t_0-t)^{-h}$ )
with the parameters
$c=1$, $h=8$,
$u_0=3$, $t_0=10$ (red line) and
$c=1$, $h=4$,
$u_0=300$, $t_0=10$  (blue line)
 (continuous line corresponds the plus sign in (\ref{pl3})  and dashed
-- "-" ).}
\end{minipage}
\hspace{10mm}
\begin{minipage}[h]{0.4\linewidth}
\includegraphics[width=7cm,height=5cm]{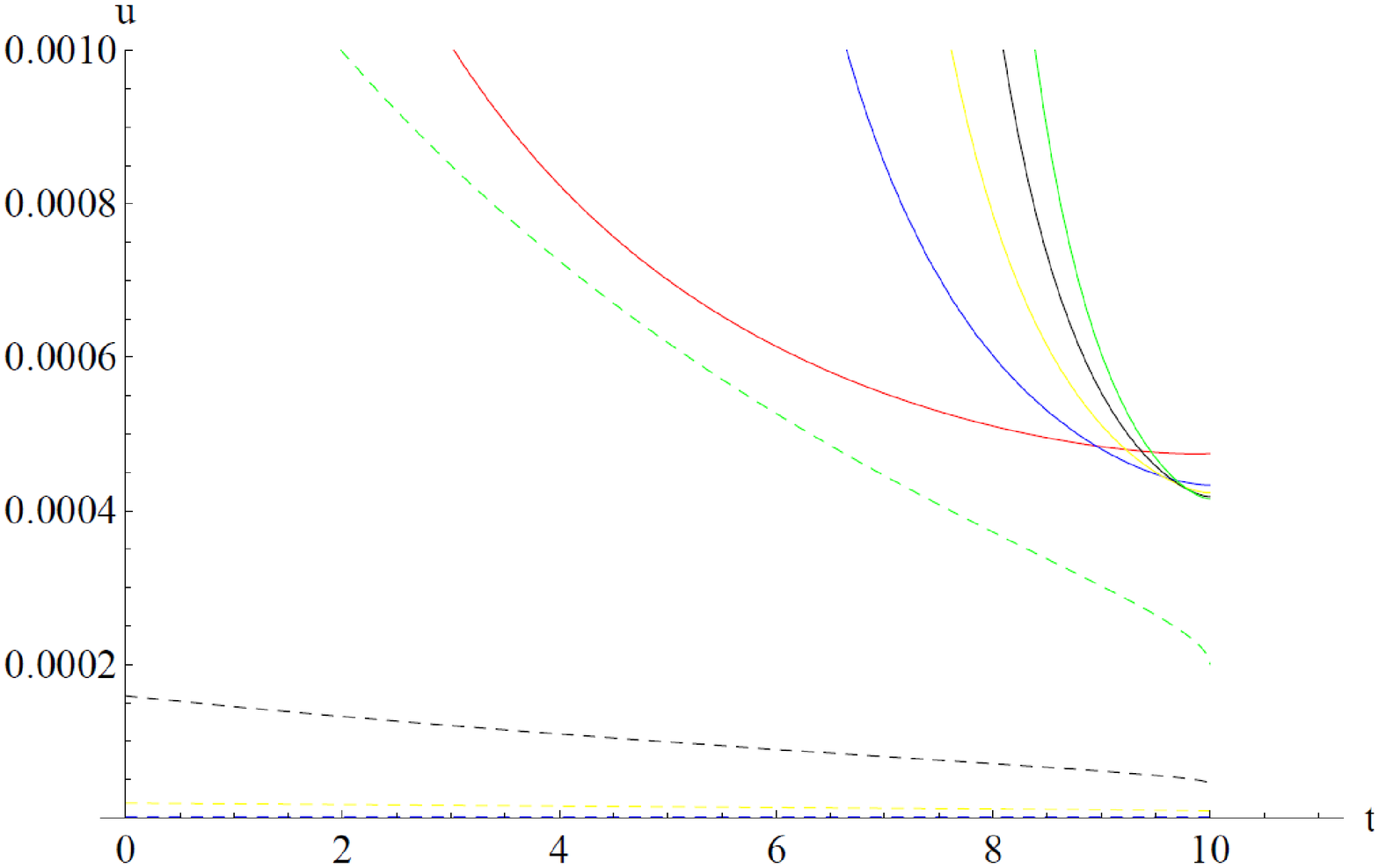}
\caption{$u$ as a function of $t$ for metric (\ref{sin11}) (continuous line corresponds $y=3/2$, dashed -- $y=1/2$)
with the parameters
$c=1$, $a_0=50$,
 $t_0=10$ $a_1 =$ 1 (red line), 3 (blue line), 5 (yellow line), 7 (black line) and 9 (green line).}
\end{minipage}
\end{figure}

In the above examples two types of singularities occur.
However, other types of finite-time future singularities are possible.
For quintessence dark energy, one can get a singularity for which the pressure
goes to infinity at a finite time, but the scale factor and density remain
finite (a sudden singularity,
or a Type II singularity) \cite{sing}. Alternatively, the density and pressure
can both become infinite with a finite scale factor
at a finite time (a Type III singularity) \cite{sing}, or higher derivatives of
the Hubble parameter $H$
can diverge (a Type IV singularity) \cite{sing}. It is known that the occurrence of a
singularity at a finite time in the future may lead to some inconsistencies.\\
Singularities of these types can be obtained by considering the metric
\be \label{sin11}
a=a_0-a_1 (t_0-t)^y,
\ee
where $a_0$, $a_1$ and $y$ are positive constants. If $y=1/2$ then we have Type
III singularity. For $y=3/2$  we have  Type II singularity. For $y=5/2$  a Type
IV singularity occurs.

Using Eq.(\ref{eeq1}) to specify the form of the scale factor one can find the
relation between the metric  $g_{\mu\nu}$ and the tensor $q_{\mu\nu}$. For example, for
$y=3/2$ we obtain the following form of $u(t)$
\begin{eqnarray}
u(t)&=&\left(36 a_0^{2/3} a_1^{4/3}\right)/\left(\left(a_0-a_1
(t_0-t)^{3/2}\right)^2 \left(-2 \sqrt{3} \text{ArcTan}\left[\frac{1+\frac{2
a_1^{1/3} \sqrt{t_0-t}}{a_0^{1/3}}}{\sqrt{3}}\right]+6 a_0^{1/3} a_1^{2/3}
c_1-\right.\right.\nonumber\\
&-&2 \left.\left. \text{Log}\left[a_0^{1/3}-a_1^{1/3}
\sqrt{t_0-t}\right]+\text{Ln}\left[a_0^{2/3}+a_0^{1/3}a_1^{1/3}
\sqrt{t_0-t}+a_1^{2/3} (t_0-t)\right]\right)^2\right),
\end{eqnarray}
where $c_1$ is  integration constant, and $c$ has been set to $c=3/4$ for simplicity.

In Fig. 4 we illustrate the behavior of the relation between the metric $g_{\mu\nu}$ and the
tensor $q_{\mu\nu}$ for $y=1/2$  ( Type III singularity) and $y=3/2$ ( Type II
singularity).  

To the light of the above results, one may ask if a $\Lambda$CDM cosmology can also be reconstructed. The answer is positive. To see it, let the scale factor have the following form
\begin{eqnarray}
a&=&a_0 e^{g(t)},\nonumber\\
g(t)&=&\frac{2}{3(1+w)} \log\left(\alpha
\,\,\text{sinh}\left(\frac{3(1+w)}{2l}(t-t_0)\right)\right),
\end{eqnarray}
where $w$, $\alpha$, $l$ and $a_0$ are constant. Such solution corresponds
  to the $\Lambda$CDM-model, which was also reconstructed in f(R) gravity\cite{LCDM}. For
this metric, the FRW equation becomes
$$\frac{3}{\kappa^2}H^2=\rho_0 a^{-3(1+w)}+\frac{3}{\kappa^2 l^2},$$
where $\alpha=\frac{1}{3}\kappa^2l^2\rho_0a_0^{-3(1+w)}$. Note, however, that when this metric is substituted into Eq.(\ref{eeq1}) (restricting the choice of the positive sign in this expression) then 
a real solution is found only if $w\leq-1/3$.  For example, if we choose $w=-1/3$ we obtain the following expression for $u$
\be
u=\frac{3 c\, l^4
\text{Csch}\left[\frac{t}{l}-\frac{t_0}{l}\right]^2}{\left(c_1-l^3
\text{Log}\left[\text{Tanh}\left[\frac{t-t_0}{2 l}\right]\right]\right)^2},
\ee
where $c_1$ is an integration constant.  The behavior of this function is illustrated in
Fig.5.\\

\begin{figure}[tbp]
\includegraphics[width=7cm,height=5cm]{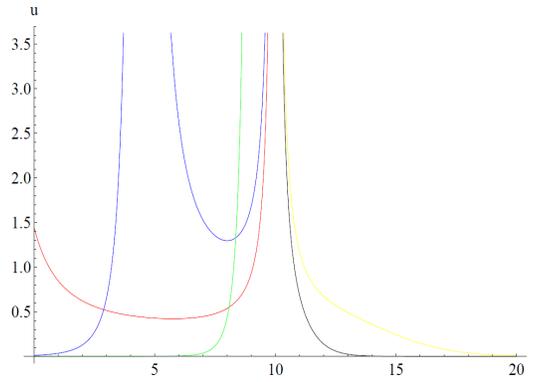}
\caption{$u$ as a function of $t$
with the parameters
$c=1$, $c_1=1$,
$t_0=10$ (red line for $l=-3$, blue for $l=-2$, green  for $l=-1$,  black  for
$l=1$ and  yellow  for $l=2$).}
\label{graph_1}
\end{figure}

We thus see that, similarly as in f(R) gravity, one can get  $\Lambda$CDM cosmology by means of modifications in the gravitational action. However, there is a qualitative difference between $f(R)$
and BI-$f(R)$ gravity in this respect. The reconstruction in $f(R)$ gravity is
achieved by the change of the function $f(R)$. In BI-$f(R)$ gravity, on the contrary, the form of $f(R)$ is uniquely fixed by internal consistency relations produced by the conformal approach. The reconstruction is possible thanks to the change of the relation between the two metrics. 
Indeed, we see that the choice of the relation between  the metric $g_{\mu\nu}$
and the tensor $q_{\mu\nu}$ in the form of the conformal connection (
$g_{\mu\nu}=u(t) q_{\mu\nu}$) uniquely determines the form of the function
$f(R)$  (\ref{ff1}) in the absence of matter or (\ref{ff2})  for matter with a
constant energy density $\rho=const$ and $p=-\rho$.
For other types of matter one should use the (more complicated and non-conformal)
Palatini formulation developed in ref.\cite{Eb6}. Hence, we obtain the
 equation (\ref{eeq1}) from which one gets either $a(t)$ or $u(t)$. It
is remarkable that one can get an arbitrary FRW cosmology as solution of
BI-$f(R)$ gravity using the above scheme.
For instance, the unification of early-time inflation with late-time
acceleration within BI-$f(R)$ gravity may be done. Of course, the corresponding
expressions are a little bit complicated and will not be presented here.

\section*{Acknowledgments}

GJO is supported by the Spanish grant FIS2011-29813-C02-02, the Consolider Program CPANPHY-1205388, the JAE-doc program and  i-LINK0780 grant of the Spanish Research Council (CSIC),  and by CNPq (Brazilian agency) through project No. 301137/2014-5.
S.D.O. and A.N.M. are  supported by
the grant of Russian Ministry of Education and Science, project TSPU-139.


\begin{thebibliography}{99}

\bibitem{review}
S. Nojiri and S.~D.~Odintsov,
   eConf C {\bf 0602061} (2006) 06
[Int.\ J.\ Geom.\ Meth.\ Mod.\ Phys.\  {\bf 4} (2007) 115] [ hep-th/0601213];
Phys.\ Rept.\  {\bf 505} (2011) 59  [arXiv:1011.0544 [gr-qc]]; Int.\ J.\ Geom.\
Meth.\ Mod.\ Phys.\  {\bf 11} (2014) 1460006
   [arXiv:1306.4426 [gr-qc]];
V.~Faraoni and S.~Capozziello,
``Beyond Einstein gravity : A Survey of gravitational theories for cosmology
and astrophysics,''
Fundamental Theories of Physics, Vol. 170, Springer, 2010;
S.~Capozziello and
M.~De Laurentis,
   Phys.\ Rept.\  {\bf 509} (2011) 167
   [arXiv:1108.6266 [gr-qc]];  G.~J.~Olmo,
  Int.\ J.\ Mod.\ Phys.\ D {\bf 20}, 413 (2011)
  [arXiv:1101.3864 [gr-qc]].


\bibitem{R}
  S.~Nojiri and S.~D.~Odintsov,
   Phys.\ Rev.\ D {\bf 68} (2003) 123512
   [hep-th/0307288].

\bibitem{BI}
S. Deser and G. W. Gibbons, Class. Quant. Grav. {\bf 15} (1998) L35;
M.
Ba\~nados and P. G. Ferreira, Phys. Rev. Lett. {\bf 105} (2010) 011101.

\bibitem{cosmology}
  X.~-L.~Du, K.~Yang, X.~-H.~Meng and Y.~-X.~Liu,
   [arXiv:1403.0083 [gr-qc]]; H.~-C.~Kim, [arXiv:1312.0703 [gr-qc]];
  P.~P.~Avelino and R.~Z.~Ferreira,Phys.\ Rev.\ D {\bf 86}  (2012) 041501
   [arXiv:1205.6676 [astro-ph.CO]];  C.~Escamilla-Rivera, M.~Banados and
P.~G.~Ferreira,Phys.\ Rev.\ D {\bf 85}  (2012) 087302
   [arXiv:1204.1691 [gr-qc]]; I.~Cho, H.~-C.~Kim and T.~Moon,
   Phys.\ Rev.\ D {\bf 86} (2012) 084018; J.~H.~C.~Scargill, M.~Banados and
P.~G.~Ferreira,  Phys.\ Rev.\ D {\bf 86} (2012) 103533; F. Fiorini, Phys. Rev.
Lett. \textbf{111} (2013) 041104; S.~I.~Kruglov,
   Phys.\ Rev.\ D {\bf 89} (2014) 064004
   [arXiv:1310.6915 [gr-qc]];K.~Yang, X.~-L.~Du and Y.~-X.~Liu,
   Phys.\ Rev.\ D {\bf 88} (2013) 124037; C.~Escamilla-Rivera, M.~Banados and
P.~G.~Ferreira, arXiv:1301.5264 [gr-qc]; T.~Harko, F.~S.~N.~Lobo, M.~K.~Mak and
S.~V.~Sushkov, Mod.\ Phys.\ Lett.\ A {\bf 29} (2014) 1450049; P.~P.~Avelino,Phys.\ Rev.\ D {\bf 85} (2012) 104053; M.~Bouhmadi-Lopez,
C.~-Y.~Chen and P.~Chen,
   [arXiv:1302.5013 [gr-qc]];  R.~Ferraro and F.~Fiorini,
   J.\ Phys.\ Conf.\ Ser.\  {\bf 314} (2011) 012114;
Dan N. Vollick,Phys.Rev. D69 (2004) 064030, [gr-qc/0309101].


\bibitem{Eb6}
A.N. Makarenko, S. Odintsov, G.J. Olmo,
[arXiv:1403.7409 [hep-th]].

\bibitem{LCDM}
S. Nojiri, S. D. Odintsov
Phys. Rev. D {\bf 74 } (2006) 086005 [hep-th/0608008];
A.~de la Cruz-Dombriz and A.~Dobado,
   Phys.\ Rev.\ D {\bf 74} (2006) 087501
   [gr-qc/0607118].

\bibitem{Eb4}
G.J. Olmo, D. Rubiera-Garcia,
  Phys.Rev. D88 (2013) 084030;
G.J. Olmo, D. Rubiera-Garcia, H. Sanchis-Alepuz, Eur. Phys. J. C  74, 2804 (2014),
[arXiv:1311.0815  [hep-th]].


\bibitem{frampton11}
P. H. Frampton, K. J. Ludwick and R. J. Scherrer, Phys. Rev. D {\bf 84} (2011)
063003  [arXiv:1106.4996 [astro-ph.CO]];
P. H. Frampton, K. J. Ludwick, S. Nojiri, S. D. Odintsov, R. J. Scherrer,
Phys. Lett. B {\bf 708} (2012) 204-211 [arXiv:1108.0067 [hep-th]]; I. Brevik, E.
Elizalde, S. Nojiri, S.D. Odintsov,
  Phys.Rev.D \textbf{84} (2011) 103508 [arXiv:1107.4642 [hep-th]];
A.~N.~Makarenko, V.~V.~Obukhov and I.~V.~Kirnos,
   {\it Astrophys.Space Sci.} {\bf 343}  (2013) 481, [arXiv:1201.4742 [gr-qc]].


\bibitem{B-R-BCS}
%

   R.~H.~Brandenberger,
   Int.\ J.\ Mod.\ Phys.\ Conf.\ Ser.\  {\bf 01} (2011) 67 [arXiv:0902.4731 [hep-th]];
   AIP Conf.\ Proc.\  {\bf 1268} (2010) 3 [arXiv:1003.1745 [hep-th]].

\bibitem{sing}
S. Nojiri, S. D. Odintsov, S. Tsujikawa,
Phys. Rev. D {\bf 71} (2005) 063004 [hep-th/0501025].



\end{thebibliography}
\end{document}